\documentclass[fleqn,twoside]{article}
\usepackage{espcrc2,amsmath,epsfig}

\newcommand{\rf}[4]{{#1} {\bf #2}, #3 (#4)}

\newcommand{\pr}{Phys.\ Rev.}

\newcommand{\physl}{Phys.\ Lett.}

\newcommand{\lapI}{$\partial^2$(I) }
\newcommand{\lapII}{$\partial^2$(II) }
\newcommand{\lapIII}{$\partial^2$(III) }

\newcommand{\oa}[1]{\ensuremath{{\cal O}(a^{#1})}}
\newcommand{\oag}[2]{\ensuremath{{\cal O}(a^{#1}g^{#2})}}

\title{Lattice Quark Propagator in Landau and Laplacian Gauges}
 
\author{Patrick O.~Bowman\address[FSU]{Department of Physics and CSIT, 
Florida State University, Tallahassee FL 32306, USA}\thanks{Presented by
POB.},
Urs M.~Heller\addressmark[FSU] and
Anthony G.~Williams\address[ADL]{CSSM and 
Department of Physics and Mathematical Physics,
Adelaide University, Australia 5005}}

\begin{document}

\begin{abstract}
We present results for the lattice quark propagator in both Landau and 
Laplacian gauges using standard and improved staggered quark actions.
The standard Kogut-Susskind action has errors of \oa{2} while the improved 
``Asqtad'' action has \oa{4}, \oag{2}{2} errors.  This improvement is seen in 
the quark propagator.  We demonstrate the application of 
tree-level corrections to these actions and see that Landau and Laplacian 
gauges produce very similar results.
\end{abstract}

\maketitle

\section{Introduction}

The quark propagator is a fundamental quantity of QCD.  Though gauge 
dependent, it manifestly displays dynamical chiral symmetry breaking, 
contains the chiral condensate and $\Lambda_{\text{QCD}}$ and has
been used to compute the running quark mass (see Ref.~\cite{Bec00b} and 
references therein).  Some model hadron
calculations rely on ans\"{a}tze for the quark propagator, yet 
on the lattice we have the opportunity to study it in a direct,
nonperturbative fashion.  Quark propagator studies can be complicated,
however, by strong lattice artefacts~\cite{Sku01a,Sku01b}.

We are required to fix a gauge and we choose the Landau and 
the Laplacian gauges~\cite{Lapgag}.  We use Wilson glue at 
$\beta = 5.85$ ($a \simeq 0.125$ fm) on a 
$16^3 \times 32$ lattice and six quark masses from 
$am = 0.075$ down to 0.0125 (115 to 19 MeV).  
Calculations were done on 80 configurations.

In the (Euclidean) continuum, Lorentz invariance allows us to decompose the 
full propagator into Dirac vector and scalar pieces
\begin{equation}
S^{-1}(p^2) = Z^{-1}(p^2) [i \gamma \cdot p + M(p^2)].
\end{equation}
Asymptotic freedom means that, as $p^2 \rightarrow \infty$, 
$S^{-1}(p^2) \rightarrow  i\gamma \cdot p + m,$ (the free propagator)
where $m$ is the bare quark mass.

We use the ``Asqtad'' quark action~\cite{Org99}, a fat-link Staggered action 
using
three-link, five-link and seven-link staples to minimise flavour changing 
interactions along with the three-link Naik
term and planar five-link Lepage term.  The 
coefficients are tadpole improved and tuned to remove all tree-level \oa{2} 
errors.  This action was motivated by the desire to improve flavour symmetry, 
but has also been reported to have good rotational properties.

From consideration of the tree-level forms of our two actions, we
define the new momentum variables $q_\mu \equiv \sin(p_\mu)$
and $q_\mu \equiv \sin(p_\mu) \bigl[ 1 + \frac{1}{6} \sin^2(p_\mu) \bigr]$
for the KS and Asqtad actions respectively.
By considering the propagator as a function of $q_\mu$, we 
ensure that the lattice quark propagator has the correct tree-level form 
and hopefully better approximates its continuum behaviour.  This is the same
philosphy that has been used in studies of the gluon propagator (see 
Ref.~\cite{Bon01} and references therein).  
See also footnote 6 in Ref.~\cite{Bec00b}.

\section{Laplacian Gauge}

Laplacian gauge is a nonlinear gauge fixing that respects rotational
invariance yet is free of Gribov ambiguities.  Although it is 
difficult to understand perturbatively, it is equivalent to Landau gauge 
in the asymptotic region.  
It is also computationally cheaper then Landau gauge.  There is, however, more
than one way of obtaining such a gauge fixing in SU(N). 
The three implementations of Laplacian gauge fixing discussed are
\begin{enumerate}
\item \lapI gauge (QR decomposition), used in Ref.~\cite{Ale01}.
\item \lapII gauge, where the complex 3x3 matrix is projected 
	onto SU(3) by maximising its trace.  This will be discussed in more
	detail in an upcoming work on the gluon propagator~\cite{Bow01b}. 
\item \lapIII gauge (Polar decomposition), the original prescription described
 	in Ref.~\cite{Lapgag}.
\end{enumerate}

The gauge transformations employed in Laplacian gauge fixing are constructed 
from the lowest eigenvectors of the covariant lattice Laplacian operator.
The three implementations discussed differ in the way that the gauge 
transformation is constructed from the above eigenvectors.  We can 
think of each projection method as defining its own Laplacian 
gauge.  In all cases the resulting gauge is unambiguous for all 
configurations except a set of measure zero.

\section{Results}

We investigated the applicaton of tree-level correction 
to the quark propagator by comparing the $Z$ functions using
$p$ and $q$.  We saw that hypercubic artefacts were small when we 
used $p$ and they vanished almost entirely when using $q$.
It is less clear which momentum variable should be used for the mass function,
because at tree-level it is not multiplied by the momentum, but for 
consistency we use $q$ here as well.  In the case of the mass 
function,  the choice of momentum will actually make little difference to our 
results.

\begin{figure}[tb]
\begin{center}
\epsfig{figure=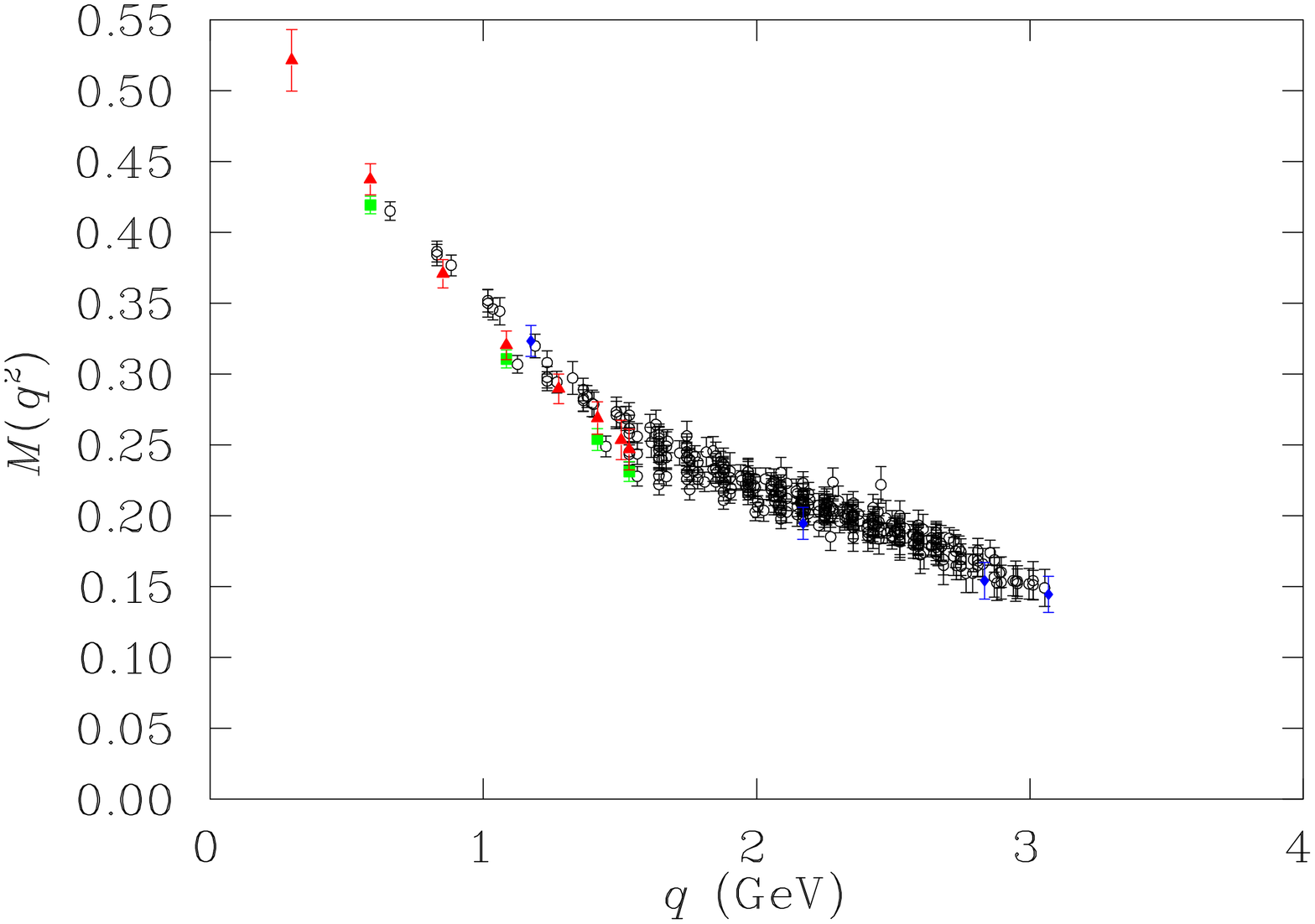,angle=90,width=7cm}
\end{center}
\begin{center}
\epsfig{figure=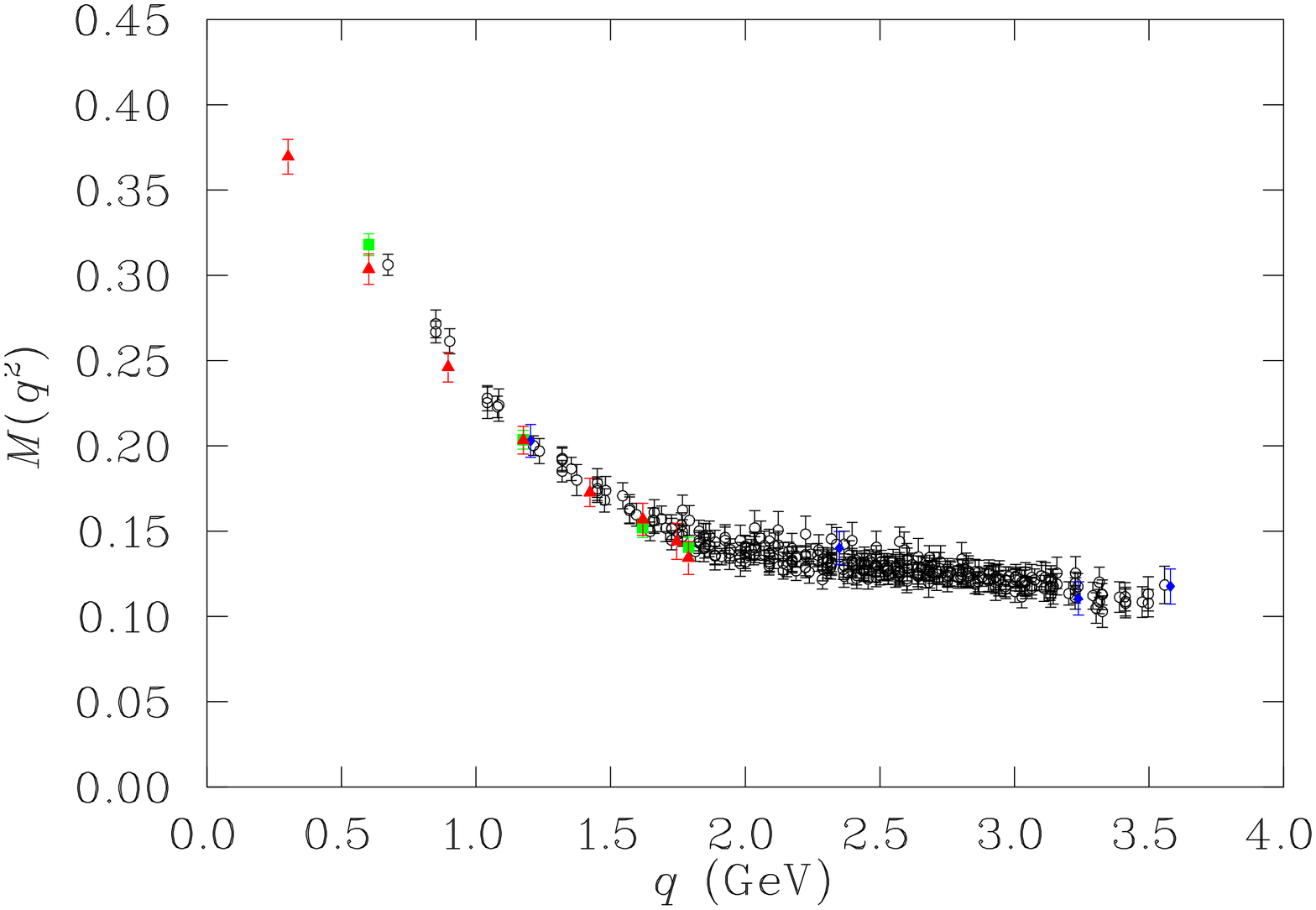,angle=90,width=7cm}
\end{center}
\vspace{-10mm}\caption{Mass function for quark mass $ma = 0.05$ 
($m \simeq 77$ MeV), KS action (top) and Asqtad action (bottom) in Landau 
gauge.}
\label{fig:lan_comp_m05all}\vspace{-5mm}
\end{figure}

In Fig.~\ref{fig:lan_comp_m05all} the mass function is plotted, in Landau
gauge, for both actions with quark mass $ma = 0.05$.  We see that the KS 
action gives a much larger value for 
M(0) than the Asqtad action and is slower to approach asymptotic behaviour.  
The Asqtad action also shows slightly better rotational symmetry.

The Asqtad action displays clearly
better rotational symmetry in the quark $Z$ function and, curiously, improved
infrared behaviour as well.  The Asqtad action also displays a better approach 
to asymptopia, approaching one in the ultraviolet.  Furthermore, the relative 
improvement increases as the quark mass decreases.  Comparing the mass 
function for the two actions at $ma = 0.0125$, the lowest mass studied here,  
the low quark mass has introduced less noise into the propagator with the 
Asqtad action than with the KS action.

\begin{figure}[bt]
\begin{center}
\epsfig{figure=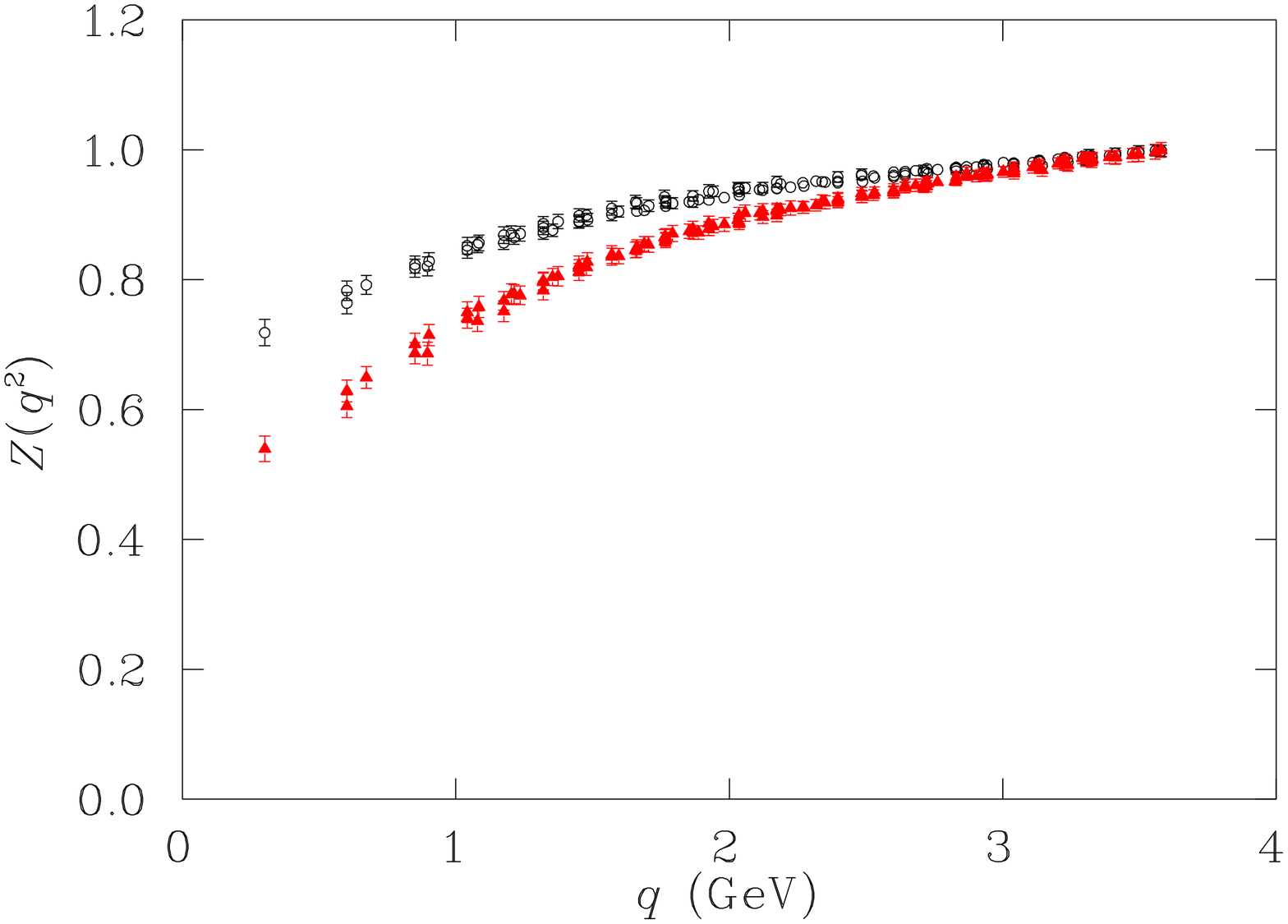,angle=90,width=7cm}
\end{center}
\begin{center}
\epsfig{figure=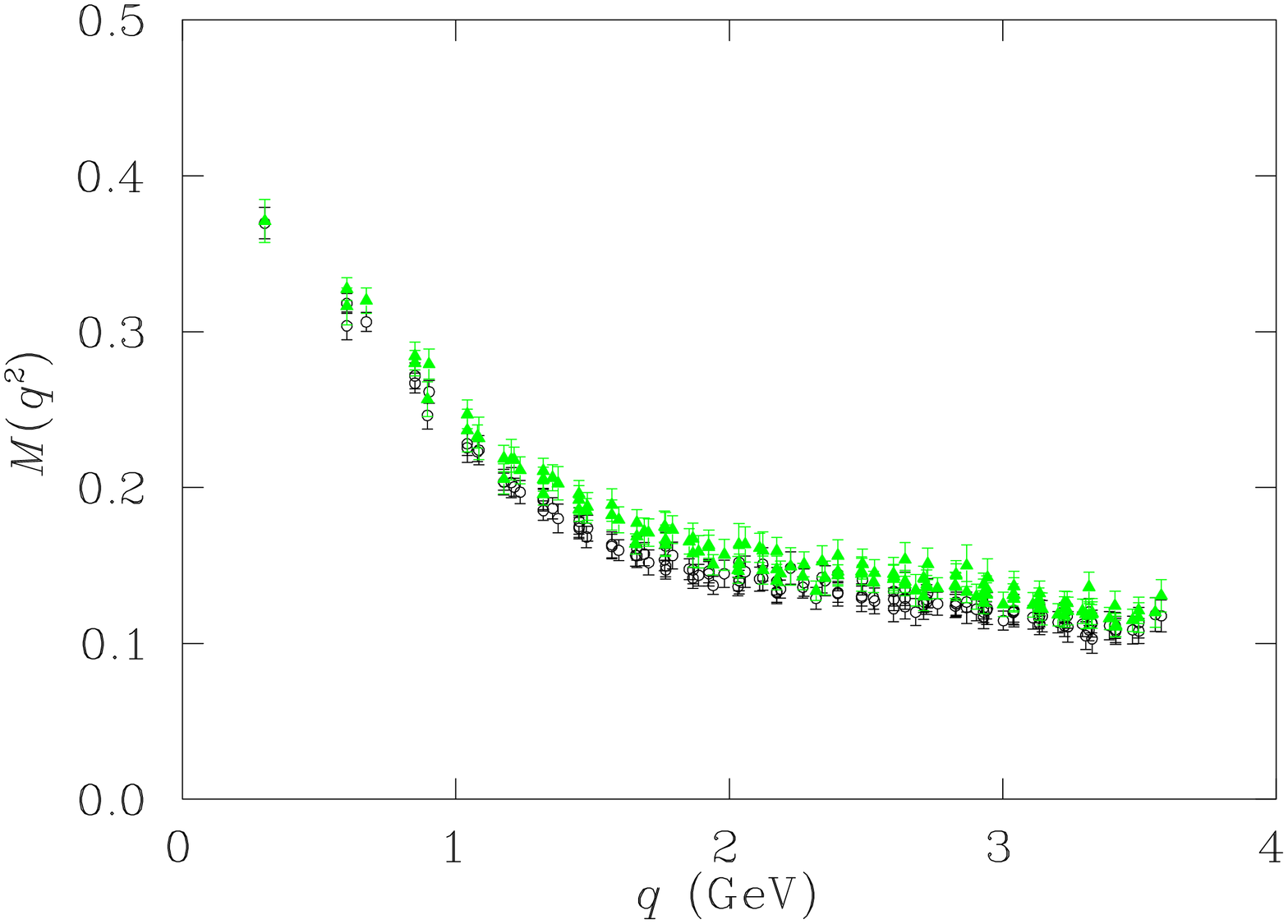,angle=90,width=7cm}
\end{center}
\vspace{-10mm}
\caption{Gauge dependence of the quark Z (top) and mass (bottom) 
functions for the Asqtad action ($ma = 0.05$).  Points marked with open 
circles are in Landau gauge and solid triangles are in \lapI gauge.  
Data has been cylinder cut.}
\label{fig:comp_asq_compm05}
\vspace{-5mm}
\end{figure}

Fig.~\ref{fig:comp_asq_compm05} (top) shows the $Z$ function for the Asqtad
action in Landau and \lapI gauges.  Data has been cylinder cut~\cite{Bon01} 
for easier comparison.They are in excellent agreement in the
ultraviolet but differ significantly in the infrared.
There appears to be some slight difference in the $Z$ function between \lapI 
and \lapII gauges.

Fig.~\ref{fig:comp_asq_compm05} (bottom) shows the mass function for the 
Asqtad action in Landau and \lapI gauges.   The two mass functions agree in 
the ultraviolet and in the infrared, but the Landau gauge mass function sits 
slightly higher in the intermediate region.  The mass functions are nearly 
identical in \lapI and \lapII gauges. 
We have also found that Landau gauge gives slightly less 
anisotropy at this lattice spacing.

\lapIII performs very poorly.  We found that many of the matrices 
had vanishingly small determinants (compared to numerical precision), which 
destroyed the projection onto SU(3). 
Problems with \lapIII have also been seen in the gluon 
propagator~\cite{Bow01b}.
 
\begin{figure}[t]
\begin{center}
\epsfig{figure=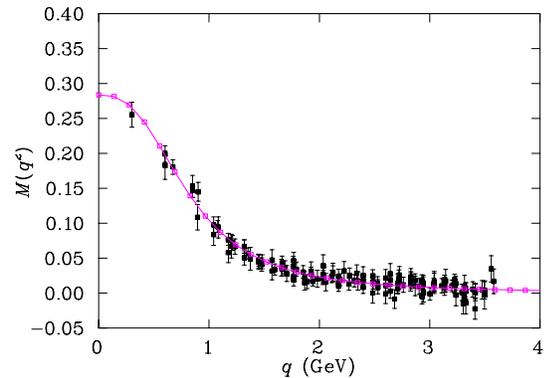,angle=90,width=7cm}
\end{center}
\vspace{-10mm}\caption{Mass function extrapolated to the chiral limit.  Errors are 
Jack-knife.  Fit parameters are
c = 0.030(4), $\Lambda$ = 870(60) MeV, $m_0$ = 0.0, $\alpha$ = 1.52(23), 
$\chi^2$ / dof = 0.49.}
\label{fig:lan_asq_chiralfit}\vspace{-5mm}
\end{figure}

A fit to each of the mass functions, including the chiral limit, was 
performed using the ansatz 
\begin{equation}
\label{eq:massfit}
M(p) = \frac{c\Lambda^{1+2\alpha}}{p^{2\alpha} + \Lambda^{2\alpha}} + m_0,
\end{equation}
which is a generalisation of the one used in Ref.~\cite{Sku01a}.  
$\alpha > 1$ is increasingly favoured as the quark mass approaches zero.
We show here, in Fig.~\ref{fig:lan_asq_chiralfit}, one fit for the mass 
function in the chiral limit.  

As we have simulated on only one lattice, it remains to do a thorough 
examination of discretisation and finite volume effects.  In forthcoming work 
we will also examine the gluon propagator in the Laplacian gauges.

\section*{Acknowledgments}

The authors wish to thank Derek Leinweber and Jonivar Skullerud for useful
discussions.


\end{document}